# Testing D-Sequences for their Randomness

Sumanth Kumar Reddy Gangasani
Oklahoma State University, Stillwater

**Abstract:** This paper examines the randomness of d-sequences, which are "decimal sequences" to an arbitrary base. Our motivation is to check their suitability for application to cryptography, spread-spectrum systems, and use as pseudorandom sequence.

## Introduction

There exist two approaches to randomness [1]: one takes the path of probability, the other that of complexity [2]. From the point of view of probability, all binary sequences of length *n* are equivalent, but from the point of view of complexity, the sequence

$$01001011\ldots110$$

is more random than the all-zero sequence

$$00000000\ldots000$$

Ritter [1] summarizes several methods to measure algorithmic complexity that includes the application of suitable transforms [3].

In this paper, we consider a new measure, due to Kak, to quantify randomness. It is based on the idea of how much the autocorrelation departs from the ideal that is true for white noise. Kak defines the randomness, *R(x)*, of a discrete sequence *x* by the expression below:

$$R(x) = 1 - \frac{\sum_{k=1}^{n-1} |C(k)|}{n-1}$$

where *C(k)* is the autocorrelation value for *k* and *n* is the period of sequence. The value of the autocorrelation is defined as in the equation below:

$$C(k) = \frac{1}{n} \sum_{j=0}^{n} a_j a_{j+k}$$

Since the randomness measure of discrete-time white noise sequence is 1 whereas that of a constant sequence is 0, this measure conforms to our intuitive expectations. Likewise, the value of randomness for a binary shift register maximum-length sequence is close to 1, in accord with our expectations.

This measure will be tried on the class of d-sequences [4-12], that are "decimal sequences" in an arbitrary base, although binary (base-2) sequences will be the only ones that will be considered in this paper. D-sequences have found several applications in cryptography, watermarking, spread-spectrum [5-12], and as generator of random numbers [13]. They constitute a very versatile source of random sequences, since, unlike maximum-length shift-register sequences [14], they are not constrained to only a limited number of period values. They have the additional virtue that other periodic sequences (including maximum-length shift register sequences) can be seen as special cases of d-sequences.

The motivation for this study is the search of new classes of random sequences for applications to cryptography and spread spectrum systems.

## Basic properties of d-sequences

We begin with a quick review of d-sequences [4-12], obtained when a number is represented in a "decimal form" in a base $r$, and it may terminate, repeat or be aperiodic. For maximum-length d-sequences of $1/q$, where $q$ is a prime number, the digits spaced half a period apart add up to $r-1$, where $r$ is the base in which the sequence is expressed.

$$\text{Ex: } \frac{1}{7} = 0.\overline{142857}$$

Here $q$ is 7, $r$ is 10, the sequence is 142857 and 6 is the period. It can be clearly seen that digits that are half the period apart add up to $r-1$ (1+8, 4+5 and 2+7).

D-sequences are known to have good cross-correlation and auto-correlation properties and they can be used in applications involving pseudorandom sequences. When the binary d-sequence is of maximum length, then bits in the second half of the period are the complements of those in the first half.

We begin with some standard results [13]:

**Theorem 1:** Any positive number x may be expressed as a decimal in the base $r$

$$A_1 A_2 ... A_{s+1} . a_1 a_2 ...$$

where $0 \le A_i \le r$, $0 \le a_i \le r$, not all $A$ and $a$ are zero, and an infinity of the $a_i$ are less than (r-1). There exists a one-to-one correspondence between the numbers and the decimals and

$$x = A_1 r^s + A_2 r^{s-1} + ... + A_{s+1} + \frac{a_1}{r} + .....$$

That the decimal sequences of rational and irrational numbers may be used to generate pseudo-random sequences is suggested by the following theorems of decimals of real numbers.

**Theorem 2:** Almost all decimals, in any base, contain all possible digits. The expression almost all implies that the property applies everywhere except to a set of measure zero.

**Theorem 3:** Almost all decimals, in any base, contain all possible sequences of any number of digits.

Theorems 2 and 3 guarantee that a decimal sequence missing any digit is exceptional. Autocorrelation and cross-correlation properties of d-sequences are given in [4] and [5].

It is easy to generate d-sequences, which makes them attractive for many engineering applications. For a maximum-length d-sequence, *1/q*, the digits of *a/q*, where *a < q*, are permutations of the digits of *1/q*, which makes it possible to use them for error-correction coding applications [5].

We are concerned primarily with maximum-length binary d-sequences (which are generated by prime numbers) because, as we will show later, the randomness measure for such sequences is generally better than for non-maximum-length d-sequences. The binary d-sequence is generated by means of the algorithm [6]:

$$a(i) = 2^i \bmod q \bmod 2$$

where *q* is a prime number. The maximum length (q-1) sequences are generated when 2 is a primitive root of *q*. When the binary d-sequence is of maximum length, then bits in the second half of the period are the complements of those in the first half.

Any periodic sequence can be represented as a generalized d-sequence *m/n*, where *m* and *n* are suitable natural numbers, i.e., positive integers.

For example, consider the maximum-length shift register [14] or PN sequence 0011101, of periods 7:

$$(.0011101)_2 = \left[\frac{1}{8} + \frac{1}{16} + \frac{1}{32} + \frac{1}{128}\right]_{10}$$

Hence $\left(\overline{.0011101}\right)_2 = \left[\frac{1}{8} + \frac{1}{16} + \frac{1}{32} + \frac{1}{128}\right] + \frac{1}{2^7}\left[\frac{1}{8} + \frac{1}{16} + \frac{1}{32} + \frac{1}{128}\right] +$

$$\frac{1}{2^{14}}\left[\frac{1}{8} + \frac{1}{16} + \frac{1}{32} + \frac{1}{128}\right] + \ldots\ldots$$

$$= \left[\frac{1}{8} + \frac{1}{16} + \frac{1}{32} + \frac{1}{128}\right]\left[1 + \frac{1}{2^7} + \frac{1}{2^{14}} + \ldots\ldots\right]$$

$$= \left[\frac{16+8+4+1}{128}\right]\left[\frac{1}{1-\frac{1}{2^7}}\right]$$

$$= \frac{29}{128}\left[\frac{128}{127}\right]$$

$$= \frac{29}{127}$$

Therefore, all PN sequences may be represented as suitable d-sequences, as shown in Figure 1.

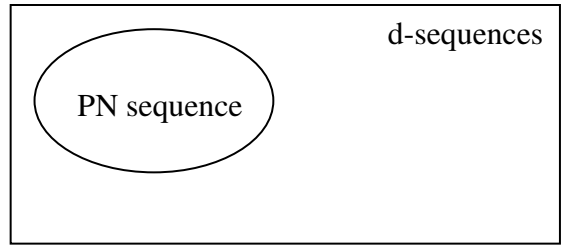

**Figure 1:** Maximum-length shift register sequences as a subset of d-sequences

## Randomness study of d-sequences

We apply the randomness measure $R(x) = 1 - \frac{\sum_{k=1}^{n-1}|C(k)|}{n-1}$, where $C(k)$ is the autocorrelation value for $k$ and $n$ is the period of sequence, to determine how it varies for different values of $q$.

We first consider the general d-sequence $1/n$, where $n$ is an odd number. For doing so we The randomness measure has its minima for $n = 2^k-1$, since for such a case the sequence would be $k$-1 0s followed by 1, which is a highly non-random sequence. However, multiplying such a sequence with an appropriate integer could generate a PN sequence as given by the previous example.

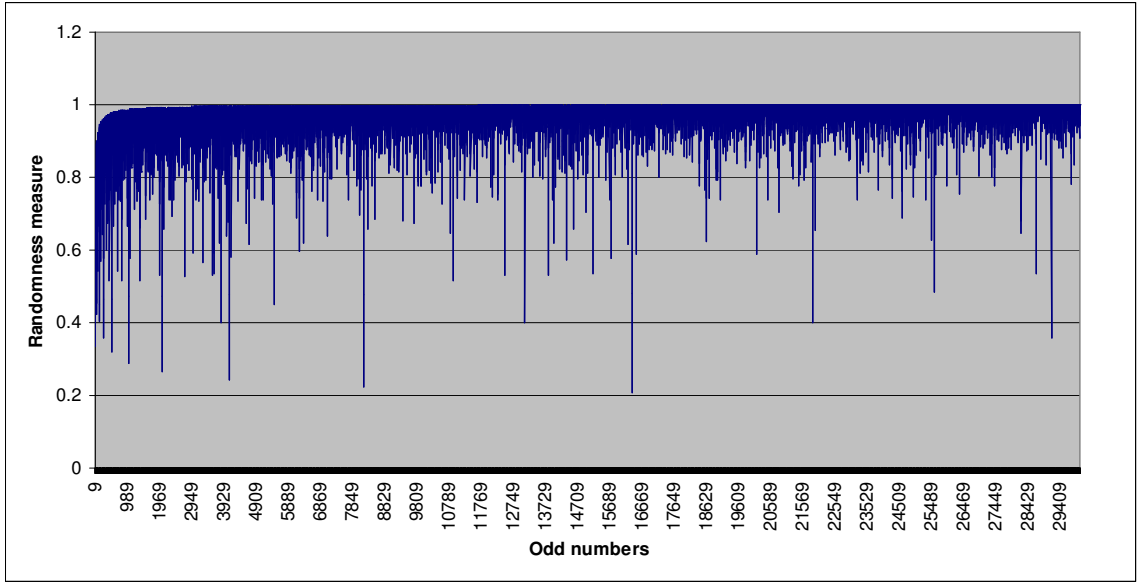

**Figure 2.** Variation of randomness measure, R, with odd numbers < 30,000

Now we consider prime numbers q < 63,180 (Figure 3).

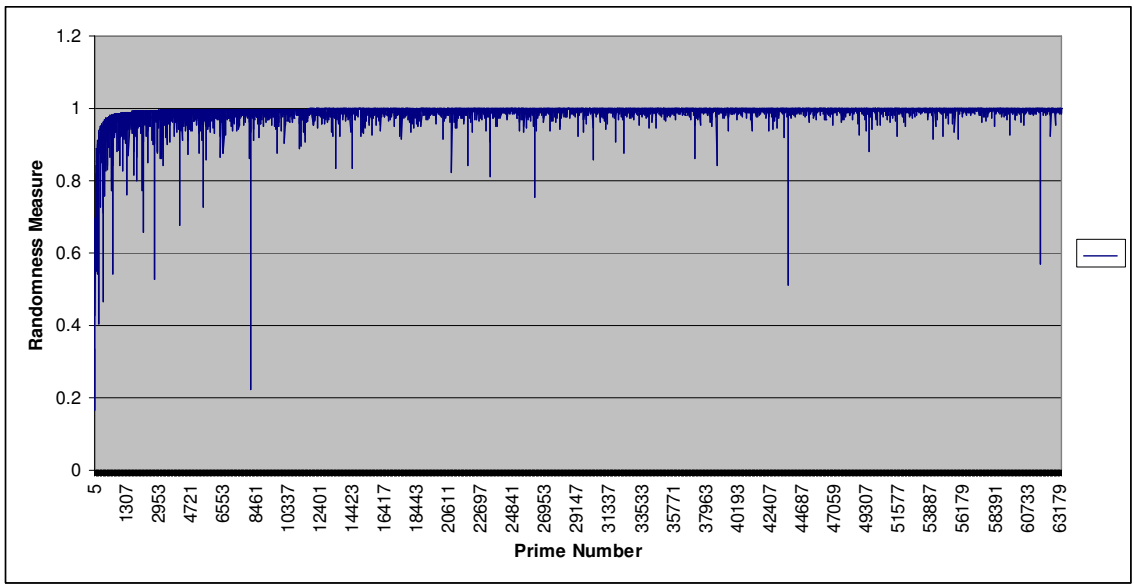

**Figure 3.** Variation of randomness measure, R, with prime numbers

We find that the randomness measure quite quickly (by q in the range of 4000 or so) climbs to a value close to 1 for maximum-length d-sequences. This increases our confidence in the use of d-sequences in cryptography applications.

Many of the primes in Figure 3 do not correspond to maximum-length sequences. As expected, the randomness measure has its minima for the Mersenne prime (8191) in Figure 3. In Figure 4, we plot the randomness measure only for maximum-length d-sequences.

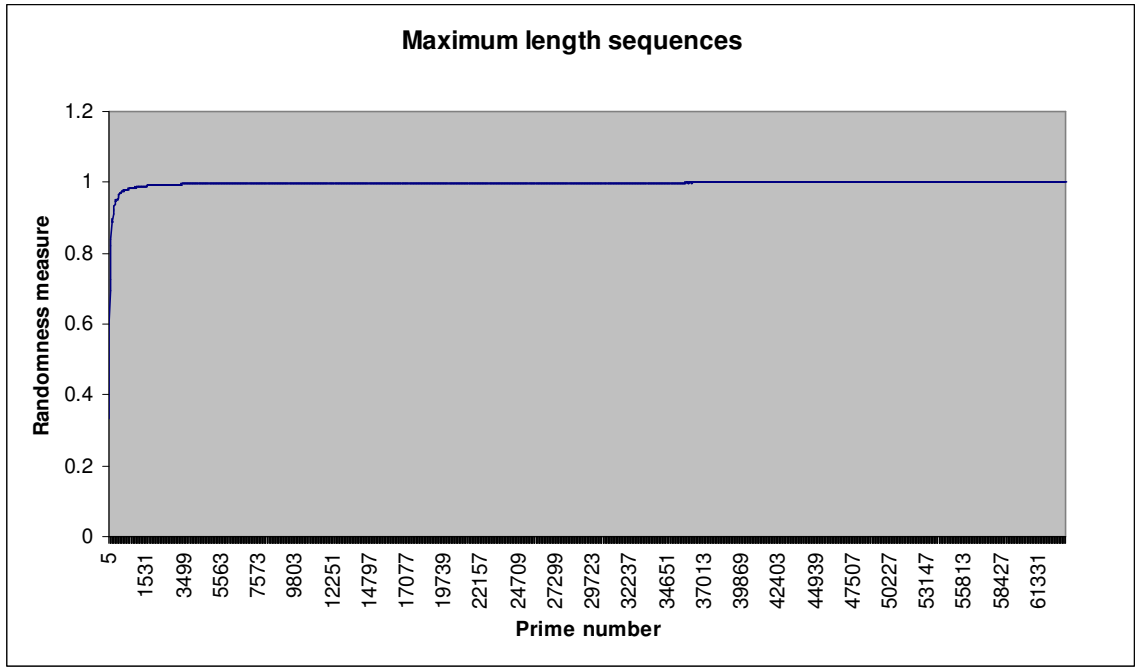

**Figure 4.** Variation of randomness measure, R, for maximum-length-sequences with prime numbers

The results are nearly equally impressive when half-length sequences are considered as in Figure 5.

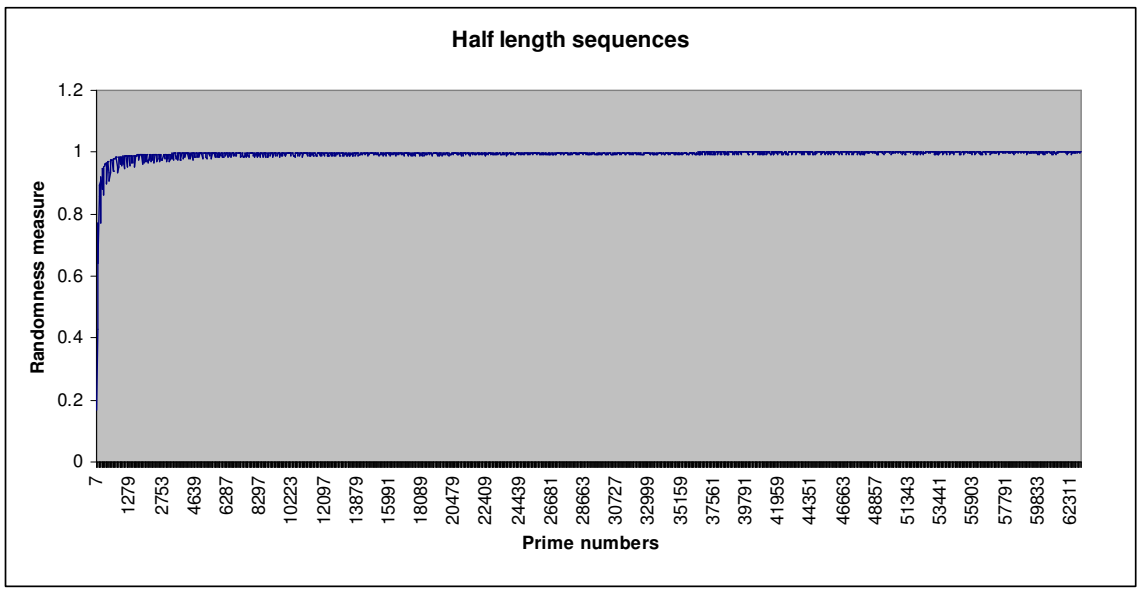

**Figure 5.** Variation of randomness measure, R, for half the maximum-length-sequences with prime numbers

## Conclusions

This note has shown that the randomness measure *R(x)* has excellent characteristics since it accords with our intuitive ideas of randomness. The application of this measure to d-sequences has shown that these have good randomness properties and, therefore, their use in cryptographic applications may be appropriate in many situations.